\documentstyle[12pt,epsf]{article}

\catcode`\@=11
\long\def\@makefntext#1{
\protect\noindent \hbox to 3.2pt {\hskip-.9pt  
$^{{\ninerm\@thefnmark}}$\hfil}#1\hfill}		

\def\@makefnmark{\hbox to 0pt{$^{\@thefnmark}$\hss}}  
	
\def\ps@myheadings{\let\@mkboth\@gobbletwo
\def\@oddhead{\hbox{}
\rightmark\hfil\ninerm\thepage}   
\def\@oddfoot{}\def\@evenhead{\ninerm\thepage\hfil
\leftmark\hbox{}}\def\@evenfoot{}
\def\sectionmark##1{}\def\subsectionmark##1{}}

\renewcommand{\thefootnote}{\fnsymbol{footnote}}

\newcounter{sectionc}\newcounter{subsectionc}\newcounter{subsubsectionc}
\renewcommand{\section}[1] {\vspace*{0.6cm}\addtocounter{sectionc}{1} 
\setcounter{subsectionc}{0}\setcounter{subsubsectionc}{0}\noindent 
	{\normalsize\bf\thesectionc. #1}\par\vspace*{0.4cm}}
\renewcommand{\subsection}[1] {\vspace*{0.6cm}\addtocounter{subsectionc}{1} 
	\setcounter{subsubsectionc}{0}\noindent 
	{\normalsize\it\thesectionc.\thesubsectionc. #1}\par\vspace*{0.4cm}}
\renewcommand{\subsubsection}[1]
{\vspace*{0.6cm}\addtocounter{subsubsectionc}{1}
      \noindent {\normalsize\rm\thesectionc.\thesubsectionc.\thesubsubsectionc. 
	#1}\par\vspace*{0.4cm}}

\newcounter{appendixc}
\newcounter{subappendixc}[appendixc]
\newcounter{subsubappendixc}[subappendixc]

\renewcommand{\appendix}[1] {\vspace*{0.6cm}
        \refstepcounter{appendixc}
        \setcounter{figure}{0}
        \setcounter{table}{0}
        \setcounter{equation}{0}
        \renewcommand{\thefigure}{\Alph{appendixc}.\arabic{figure}}
        \renewcommand{\thetable}{\Alph{appendixc}.\arabic{table}}
        \renewcommand{\theappendixc}{\Alph{appendixc}}
        \renewcommand{\theequation}{\Alph{appendixc}.\arabic{equation}}
        \noindent{\bf Appendix \theappendixc #1}\par\vspace*{0.4cm}}

\def\abstracts#1{{
 \centering{\begin{minipage}{12.2truecm}\footnotesize\baselineskip=12pt\noindent
	\centerline{\footnotesize ABSTRACT}\vspace*{0.3cm}
	\parindent=0pt #1
	\end{minipage}}\par}} 

\renewenvironment{thebibliography}[1]
	{\begin{list}{\arabic{enumi}.}
	{\usecounter{enumi}\setlength{\parsep}{0pt}
\setlength{\leftmargin 1.25cm}{\rightmargin 0pt}
	 \setlength{\itemsep}{0pt} \settowidth
	{\labelwidth}{#1.}\sloppy}}{\end{list}}

\newcounter{itemlistc}
\newcounter{romanlistc}
\newcounter{alphlistc}
\newcounter{arabiclistc}

\newcommand{\fcaption}[1]{
        \refstepcounter{figure}
        \setbox\@tempboxa = \hbox{\footnotesize Fig.~\thefigure. #1}
        \ifdim \wd\@tempboxa > 6in
           {\begin{center}
        \parbox{6in}{\footnotesize\baselineskip=12pt Fig.~\thefigure. #1}
            \end{center}}
        \else
             {\begin{center}
             {\footnotesize Fig.~\thefigure. #1}
              \end{center}}
        \fi}

\newcommand{\tcaption}[1]{
        \refstepcounter{table}
        \setbox\@tempboxa = \hbox{\footnotesize Table~\thetable. #1}
        \ifdim \wd\@tempboxa > 6in
           {\begin{center}
        \parbox{6in}{\footnotesize\baselineskip=12pt Table~\thetable. #1}
            \end{center}}
        \else
             {\begin{center}
             {\footnotesize Table~\thetable. #1}
              \end{center}}
        \fi}

\def\@citex[#1]#2{\if@filesw\immediate\write\@auxout
	{\string\citation{#2}}\fi
\def\@citea{}\@cite{\@for\@citeb:=#2\do
	{\@citea\def\@citea{,}\@ifundefined
	{b@\@citeb}{{\bf ?}\@warning
	{Citation `\@citeb' on page \thepage \space undefined}}
	{\csname b@\@citeb\endcsname}}}{#1}}

\newif\if@cghi
\def\cite{\@cghitrue\@ifnextchar [{\@tempswatrue
	\@citex}{\@tempswafalse\@citex[]}}
\def\citelow{\@cghifalse\@ifnextchar [{\@tempswatrue
	\@citex}{\@tempswafalse\@citex[]}}
\def\@cite#1#2{{$\null^{#1}$\if@tempswa\typeout
	{IJCGA warning: optional citation argument 
	ignored: `#2'} \fi}}

 1
 1
 1

\font\ninerm=cmr9

\newcommand{\ra}{\rightarrow}
\newcommand{\gev}{\,GeV}
\newcommand{\tev}{\,TeV}
\newcommand{\pbinv}{\,pb^{-1}}
\newcommand{\fbinv}{\,fb^{-1}}
\newcommand{\beq}{\begin{equation}}
\newcommand{\eeq}{\end{equation}}
\begin{document}

\centerline{\normalsize\bf $h^0\rightarrow W^+W^-\ra \ell^+\ell^{'-}\nu_\ell{
\bar \nu}_{\ell'}$ as the Dominant SM Higgs} 
\baselineskip=22pt
\centerline{\normalsize\bf Search Mode at the LHC for $M_{h^0}= 155-180\gev$}

\centerline{\footnotesize Michael Dittmar}
\baselineskip=13pt
\centerline{\footnotesize\it ETH-Z\"urich} 
\baselineskip=12pt 
\centerline{\footnotesize\it CH-8093 Z\"urich }
\centerline{\footnotesize E-mail: Michael.Dittmar@cern.ch}
\vspace*{0.3cm}
\centerline{\footnotesize and}
\vspace*{0.3cm}
\centerline{\footnotesize Herbi Dreiner}
\baselineskip=13pt
\centerline{\footnotesize\it Rutherford Laboratory}
\baselineskip=12pt 
\centerline{\footnotesize\it Chilton, Didcot, Oxon, OX11 0QX}
\centerline{\footnotesize E-mail: dreiner@v2.rl.ac.uk}

\vspace*{0.9cm}
\abstracts{We show that the decay $h^0\ra W^+W ^-\ra\ell^+\ell^-\nu_\ell{
\bar\nu}_\ell$ is the most sensitive mode for SM Higgs searches in the range
$155-180\,GeV$. The previously considered mode $h^0\ra Z^0Z^{0*}\ra\ell^+\ell^
-\ell^+\ell^-$ has a significantly lower search sensitivity. We place
particular emphasis on two new cuts based on (i) the boost and (ii) the
spin-correlation of the $W^+W^-$-system. The distribution we obtain from our
combined cuts shows a mass sensitive peak which probably allows a mass 
determination to $\pm5\gev$ for $5\fbinv$. This contribution complements our
paper.\cite{paper} }

\normalsize\baselineskip=15pt
\setcounter{footnote}{0}
\renewcommand{\thefootnote}{\alph{footnote}}
\section{Introduction}
\noindent The Higgs boson is the missing building block of the Standard Model
It is imperative to find it as soon as possible. The LHC will initially
have an integrated luminosity of $10\fbinv/y$, upgraded to $100\fbinv/y$
several years later. The lower luminosity will leave a gap in search
sensitivity for Higgs masses between\cite{denegri} $155-180\,GeV$. The
previously considered mode $h^0 \ra Z^0Z^{0*}$ requires at least
$100\,fb^{-1}/y$. Thus we might not find the Higgs boson for several years {\it
after} the LHC has been running. Here, we show how to fill
this gap by the decay\cite{paper} 
\beq 
h^0\ra W^+W^-\ra \ell^+\nu_\ell\ell^{'-}{\bar\nu}_{\ell'}.
\label{eq:signal}
\eeq 
This leads to $10^3$ times as many events as $h^0\ra Z^0(Z^0)^*\ra \ell^+
\ell^-\ell^{'+}\ell^{'-}$ which can compensate for the lack of a
reconstructed narrow mass peak. Furthermore, an integrated luminosity of only $5
\fbinv$ is sufficient for discovery! Over the entire Higgs mass range accessible
to the LHC this is the smallest required luminosity of any search mode. The
final distribution is sensitive to the Higgs mass probably allowing a
determination to $\pm5\gev$ for $5\fbinv$.

\section{Separating the Signal from the Continuum $WW$ Background}
\noindent The signature (\ref{eq:signal}) has been studied in two parton level
analyses\cite{nigel,vernon}. Both were modestly optimistic for the LHC but were
subsequently ignored. We go beyond these analyses\cite{paper} in several
respects. {\bf (1)} Most importantly, in all cases a full simulation of QCD
processes, including hadronisation processes is done using the PYTHIA Monte
Carlo\cite{pythia}. All $K$-factors are set to one since the full set is not
yet known for the background processes. Their inclusion would most likely
improve the significance of the search since for the dominant Higgs production
process\cite{spira} it is known to be large. {\bf (2)} We include the leptonic
decays of the tau lepton. This slightly degrades the signal. {\bf (3)} We
include all possible background processes with at least two isolated leptons,
including $gg\ra Wtb$, which is of the same order as $t{\bar t}$ pair
production after the initial set of cuts. {\bf (4)} When including the above,
in particular {\bf (1)}, the old cuts\cite{nigel,vernon} are no longer
sufficient. The two main new cuts we impose are based on the boost and the spin
correlation of the WW-system. We explain these in detail below. 

The signal cross section is $1.2\pbinv$ including the leptonic branching
ratios. The two main background processes and their cross sections multiplied
by the leptonic decay BR's are  
\begin{eqnarray}
q{\bar q}&\ra& W^+W^-\ra\ell^+\nu_\ell\ell^{'-}\nu_{\ell'}, \quad7.4\pbinv,
\label{eq:qqww}\\ 
q{\bar q},gg&\ra& t{\bar t}\ra W^+W^-b{\bar b}, \quad\;\;\;\;\;\;\;\;\;\,
62.5\pbinv. 
\end{eqnarray}
The signal is most difficult to distinguish from the irreducible background
(\ref{eq:qqww}). The central events of the $t{\bar t}$ background are 
controlled by the jet rejection cut below. 

We first implement a set of cuts which are fairly standard. {\it (i)} We require
two identified charged leptons with opposite charge. {\it (ii)} The $p_t$ of the
leptons should be greater than $10\gev$, and the absolute value of their 
rapidity $|\eta|$ should be less than 2. Furthermore, in a cone with half-angle 
$20^o$ around each lepton the hadronic and electromagnetic energy should be less
than $5\gev$. These first two criteria select events with two isolated leptons
according to ATLAS/CMS capabilities. {\it (iii)} For one lepton $p_t>20\gev$. {
\it (iv)} We require the absence of any jet with a $p_t>20\gev$ and $|\eta_{
jet}|<2.4$. {\it (v)} The dilepton invariant 
mass $M_{\ell\ell}<80\gev$. {\it (vi)}
The missing transverse momentum of the dilepton system $\not\!\!{p_t}(\ell\ell
)>20\gev$. {\it (vii)} In the plane transverse to the beam we require the angle
between the two charged leptons $\Delta\phi_\perp(\ell\ell)<135^o$. These
cuts combined select events with $W^+W^-+X.$
\begin{figure}[t]
   \vspace{-1.0cm}
   \epsfysize=15cm
   \epsfxsize=14cm
   \centerline{\epsffile{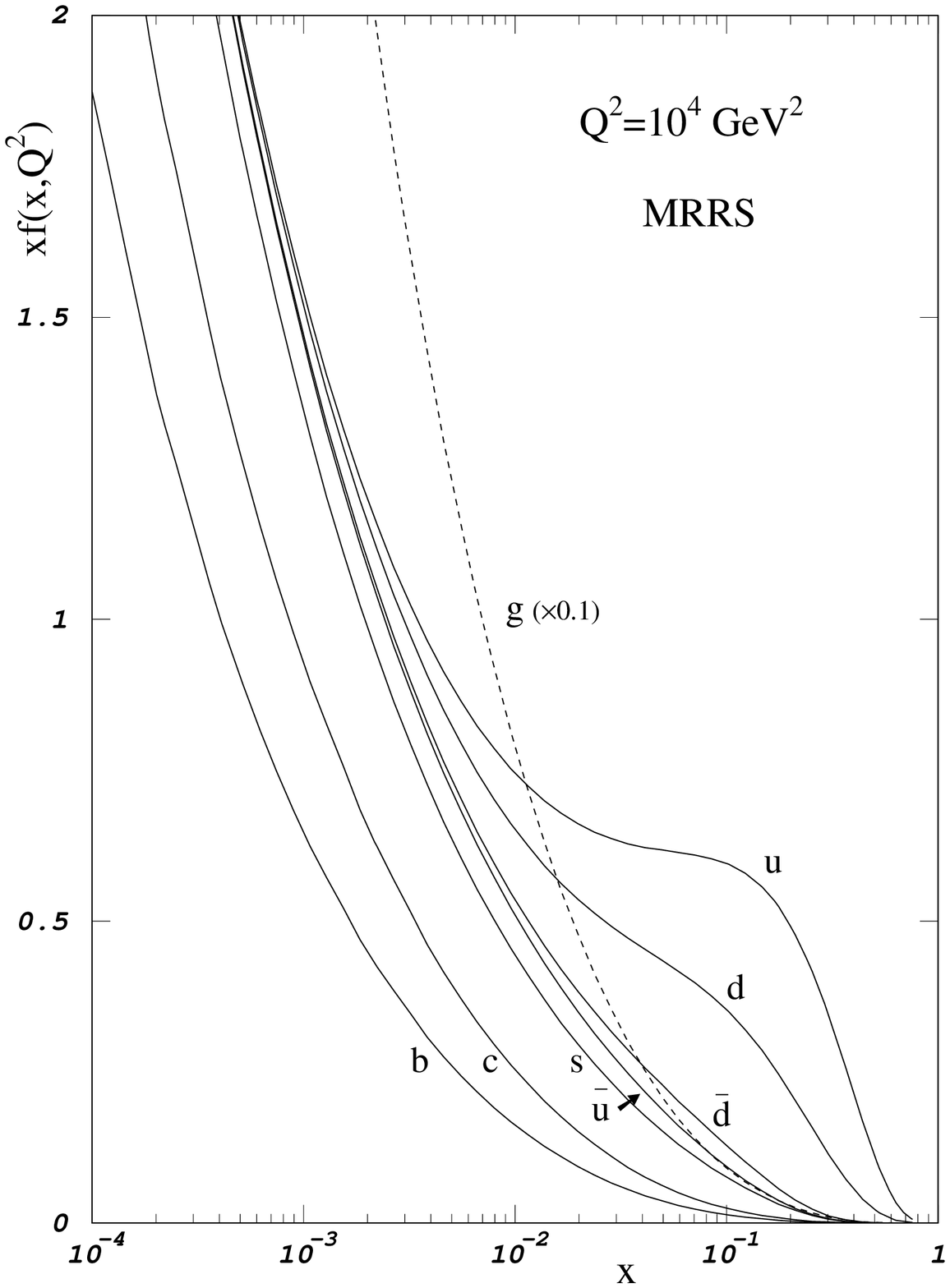}}
   \vspace*{-0.6cm}
\caption[dummy]{\label{struc} \small 
MRRS\cite{mrrs} structure functions scaled by
$x$. $g(x)$ is scaled by $x/10$.} 
   \vspace*{-0.5cm}
\end{figure}
The most serious background is (\ref{eq:qqww}). Compared only to this we obtain
the significance for $M_{h^0}=170 \gev$: $S/\sqrt{B_{WW}}=13$, and a signal to
background ratio of $S/B_{WW}=1/3$. For the full set of backgrounds we obtain
$S/\sqrt{B}=7$, and $S/B=1/10$. This requires an understanding of the
the background to better than $2\%$, which is perhaps not realistic. We aim to
improve this with the following cuts which specifically attack the
$WW$-continuum background. 

\begin{figure}[t]
   \vspace{-2.0cm}
   \epsfysize=15cm
   \epsfxsize=14cm
   \centerline{\epsffile{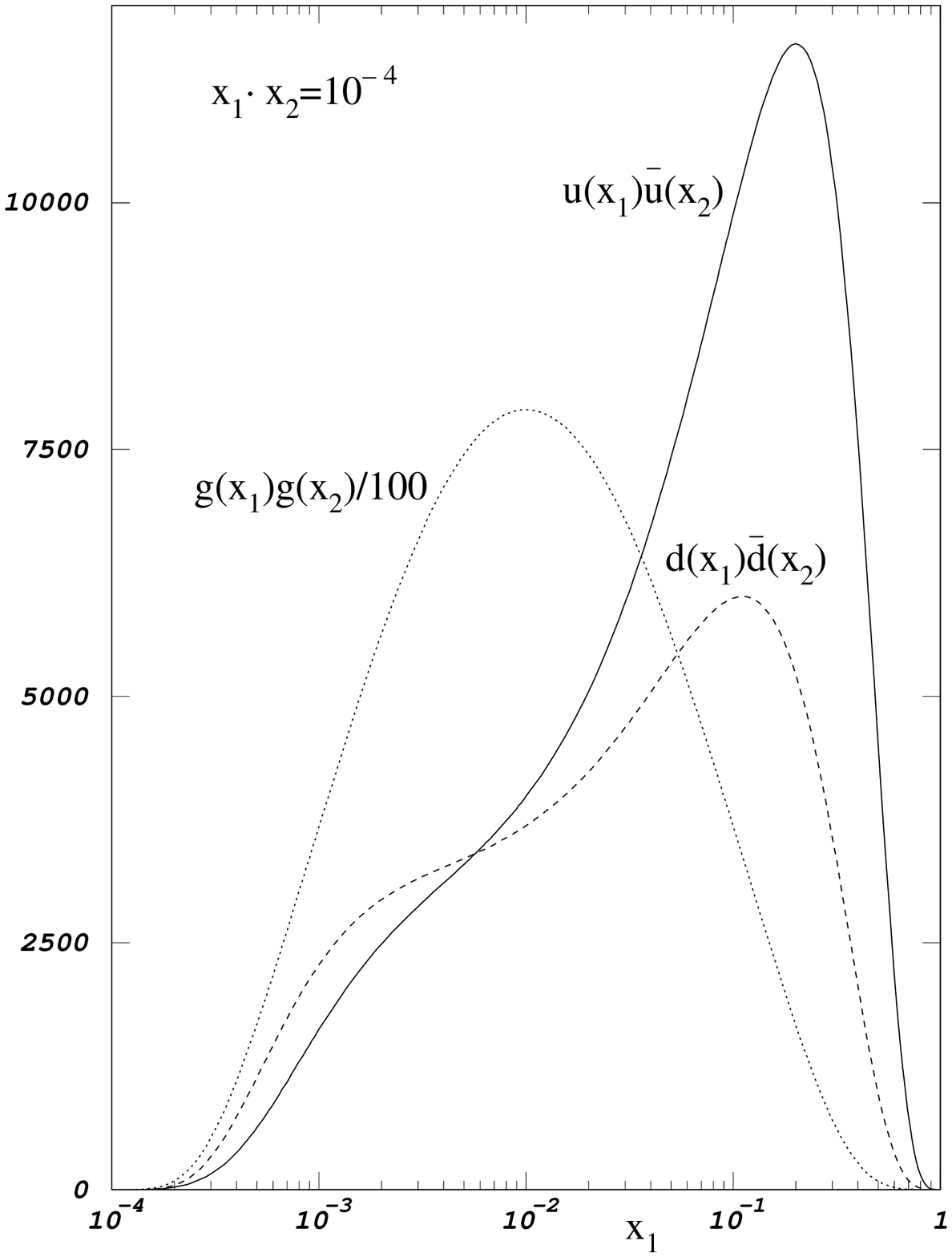}}
   \vspace*{-0.6cm}
\caption[dummy]{\label{strucprod} \small Product of the structure functions as
a function of $x_1$ for fixed $x_1\cdot x_2=10^{-4}$.} 
   \vspace*{-0.5cm}
\end{figure}
\subsection{Boost of the $W^+W^-$-System}
\noindent Consider $q{\bar q}\ra W^+W^-$ where $q$ has a momentum fraction
$x_1$ of one proton and ${\bar q}$ a momentum fraction $x_ 2$ of the other
proton.\footnote{$g+g\ra WW$ is a potential further important background 
process. It is one-loop suppressed but is enhanced by the gluon luminosity and
the coherent sum over the quark flavours giving an overall factor of about
3600. All the same, this contribution is almost an order of magnitude
less\cite{dicus} than (\ref{eq:qqww}). It thus does not affect our analysis
significantly and we have omitted it. (It is also not yet included in PYTHIA.)
However, it should be included in a complete analysis of this problem.} Then 
\beq 
x_1\cdot x_2 = \frac{{\hat s}}{s}=\frac{M^2_{WW}}{s}\approx 10^{-4}.
\label{eq:energy}
\eeq 
Here ${\hat s}$ is the parton-level center-of-mass energy squared. $\sqrt{s}$ is
the centre-of-mass energy of the LHC, $14\,TeV$, and $M_{WW}$ is the invariant 
mass of the $WW$-system. For the signal, $M_{WW}$ is equal to the Higgs mass. 
For the background process $M_{WW}\approx 155-180\,GeV$, giving 
(\ref{eq:energy}).

The total background production cross section is given by the parton level cross
section folded with the product of the parton distribution functions $q(x_1)
\cdot{\bar q}(x_2)$. In Figure 1 we show the individual parton distribution
functions $x\cdot q(x)$, $x\cdot g(x)$. In Figure 2 we show the products which
contribute to the signal and background. Here we have fixed $x_1\cdot x_2=10^
{-4}$. We see that both $u{\bar u}$ and $d{
\bar d}$ strongly peak at about $0.1$, well away from $0.01$. In contrast $gg$
has a broad peak at about $0.01$. This is because, as seen in Figure 1, both
$xu(x)$ and $xd(x)$ have a clear shoulder at about $0.1$ due to their valence
quark distributions and $xg(x)$ shows no such shoulder.
\begin{figure}[t]
   \vspace{-2.0cm}
   \epsfysize=11.5cm
   \epsfxsize=10cm
   \centerline{\epsffile{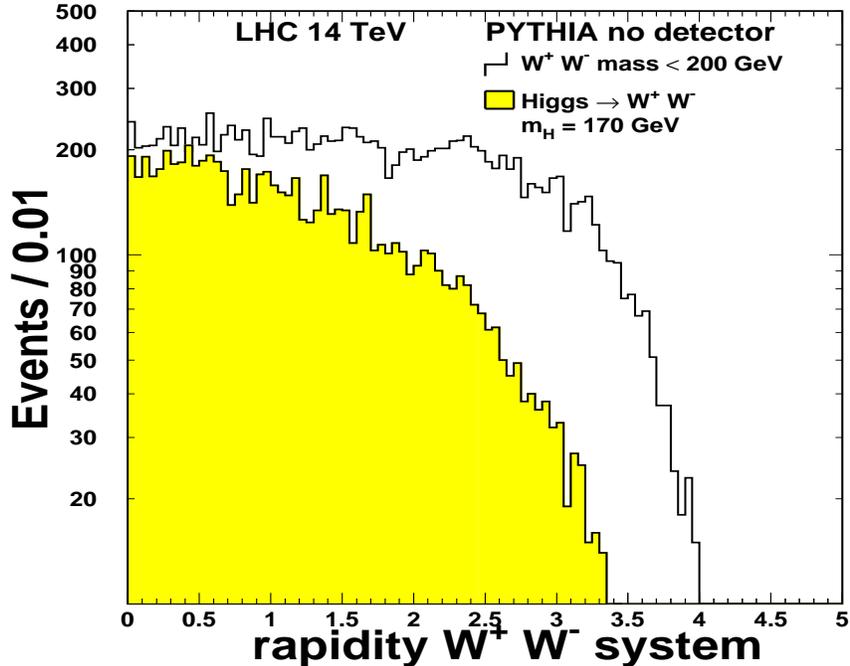}}
   \vspace*{-1.9cm}
\caption[dummy]{\label{rapidity} Monte Carlo generated distribution of
the signal (shaded histogram) and background $WW$ rapidity distributions.}
   \vspace*{-0.5cm}
\end{figure}

At the peak of $u{\bar u}$, $x_1=0.2\gg x_2=5\,10^{-4}$. Due to this large
momentum imbalance, the Lorentz boost of the $WW $-system is $\gamma=(x_1+x_2)/
(\sqrt{4x_1x_2})\approx10$. This corresponds to a momentum for the $WW$-system
of about $1.4\tev\gg M_W$. The dominant signal process\cite{moretti} is $gg\ra
h^0\ra WW$. From Figure 2 it is clear that this has a much lower boost. Indeed,
at the peak value of $gg$ the boost vanishes. This is confirmed by Figure 3
where we show the raw ({\it i.e.} without cuts {\it (i)-(vii)}) Monte Carlo
generated $WW$-rapidity distributions of both the signal and the continuum
background. The background has a plateau extending to about 2.5-3 in rapidity
whereas the signal has fallen off by a factor of 3.

Of course, we can not observe the WW-system, only the charged leptons. Due to 
the tremendous boost, the decay leptons for the background will be strongly
folded towards the beam axis, while the signal leptons will be central. We thus 
require $\cos\theta_{\ell\ell}<0.8$, where $\theta_{\ell\ell}$ is the angle of
the dilepton system with respect to the beam.

\subsection{Spin Correlation of the $W^+W^--$System}
\noindent The Higgs boson has spin zero, the $W^\pm$ bosons spin 1. In order to
conserve angular momentum, the spins of the W-bosons from $h^0\ra WW$ must be
anti-correlated. In the Higgs rest-frame (which for the considered mass range is
practically the lab frame),
we denote the decay axis of the the $WW$-system the $3$-axis. Along this
axis, the W-spins are quantized $S_3(W)=\pm1,0$. These are denoted transverse
($T$) and longitudinal ($L$), respectively. Thus only the decays 
\begin{eqnarray}
h^0\;\ra\; W_T^+W_T^-,\quad h^0\;\ra\; W_L^+W_L^-
\end{eqnarray}
are allowed, whereas $h^0\rightarrow W_L^\pm W_T^\mp$ is prohibited. 

The $W^\pm$ polarizations are not directly observable, instead we observe the
final state charged leptons. The decay rate of ${\vec{W}}^+_T\ra e^++\nu_e$ is
proportional to $(1+\cos\vartheta)^2$, where $\vartheta$ is the angle of the
positron ${\vec{{\bf p}}}_{e^+}$ with respect to the $W^+_T$ spin. Thus the
(right-handed) positron is preferentially emitted in the {\it same} direction
as the $W^+_T$ spin. Analogously, the (left-handed) electron is emitted in the
{\it opposite} direction of the $W^-_T$-spin with a $(1-\cos \vartheta)^2$
distribution. Since the W-spins are {\it anti}-correlated, ${\vec{{\bf p}}}_{e^
+}$, ${\vec{{\bf p}}}_{e^-}$ are in the {\it same} direction. 

\begin{figure}[t]
   \vspace{-1.0cm}
   \epsfysize=14cm
   \epsfxsize=13cm
   \centerline{\epsffile{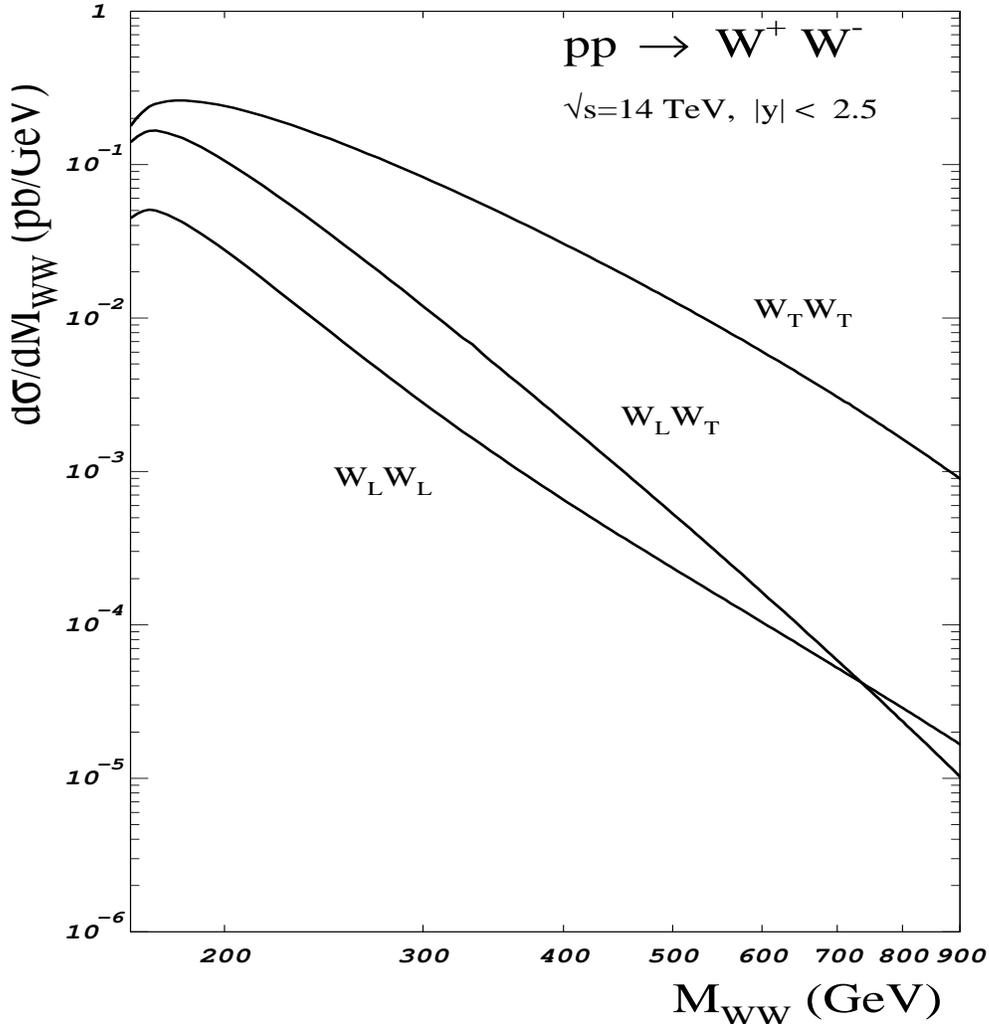}}
   \vspace*{-0.2cm}
\caption[dummy]{\label{dsdmm} \small Invariant mass distribution for $W$ boson
pair production from $q{\bar q}$ annihilation at the LHC. L (longitudinal) and
T (transverse) refer to the polarization of the $W$ bosons. The differential
cross section in Ref\cite{scott} was used.} 
  \vspace*{-0.2cm}
\end{figure}

The charged lepton from the decay of a $W_L^\pm$ has a $\sin^2\vartheta$
distribution, where $\vartheta$ is the angle of ${\vec{{\bf p}}}{e}$ with
respect to the $3$-axis. The lepton is most likely to be emitted perpendicular
to the $3$-axis. If the $W$-boson decays were uncorrelated there would be no
particular correlation between ${\vec{{\bf p}}}_{e^+}$ and ${\vec{{\bf p}}}_{e^
-}$. However, eventhough the $W$'s may decay outside of eachother's lightcone
their decays are correlated: "they know of eachother". \footnote{This is 
analogous to the correlated measurements of photon spins in tests of Bell's
inequality.} The correlated decay-rate can be calculated\cite{nigel} and is
proportional to $(e^-\cdot\nu_e)(e^+\cdot{\bar\nu}_e)$, where we have denoted
the 4-momenta by the corresponding particle symbol. This is zero for ${\vec{{
\bf p}}}_{e^+}$, ${\vec{{\bf p}}}_{e^-}$ anti-parallel and is maximum for them 
being parallel, just as in the $W_T^-W_T^+$ case. Overall, we thus expect for 
the signal that ${\vec{{\bf p}}}_{e^+}$, ${\vec{{\bf p}}}_{e^-}$ have a small 
relative opening angle. 

\begin{figure}[t]
   \vspace{-2.0cm}
   \epsfysize=15.0cm
   \epsfxsize=14cm
   \centerline{\epsffile{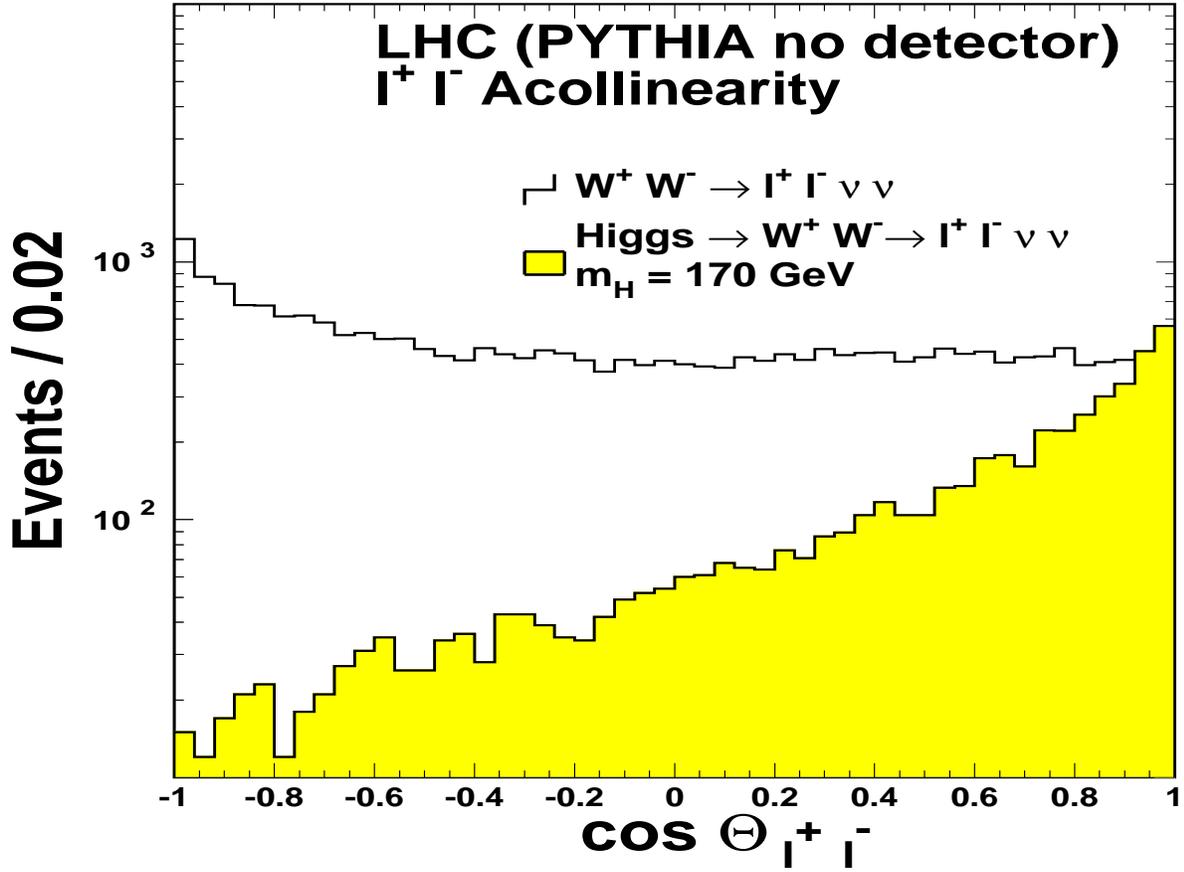}}
   \vspace*{-1.7cm}
\caption[dummy]{\label{acolin} \small Signal (shaded histogram) and background 
distribution of events in $\cos\theta_{\ell^+\ell^-}$ for $5\fbinv$ integrated 
luminosity.} 
   \vspace*{-0.2cm}
\end{figure}
\begin{figure}[t]
   \vspace{-2.0cm}
   \epsfysize=15.0cm
   \epsfxsize=14cm
   \centerline{\epsffile{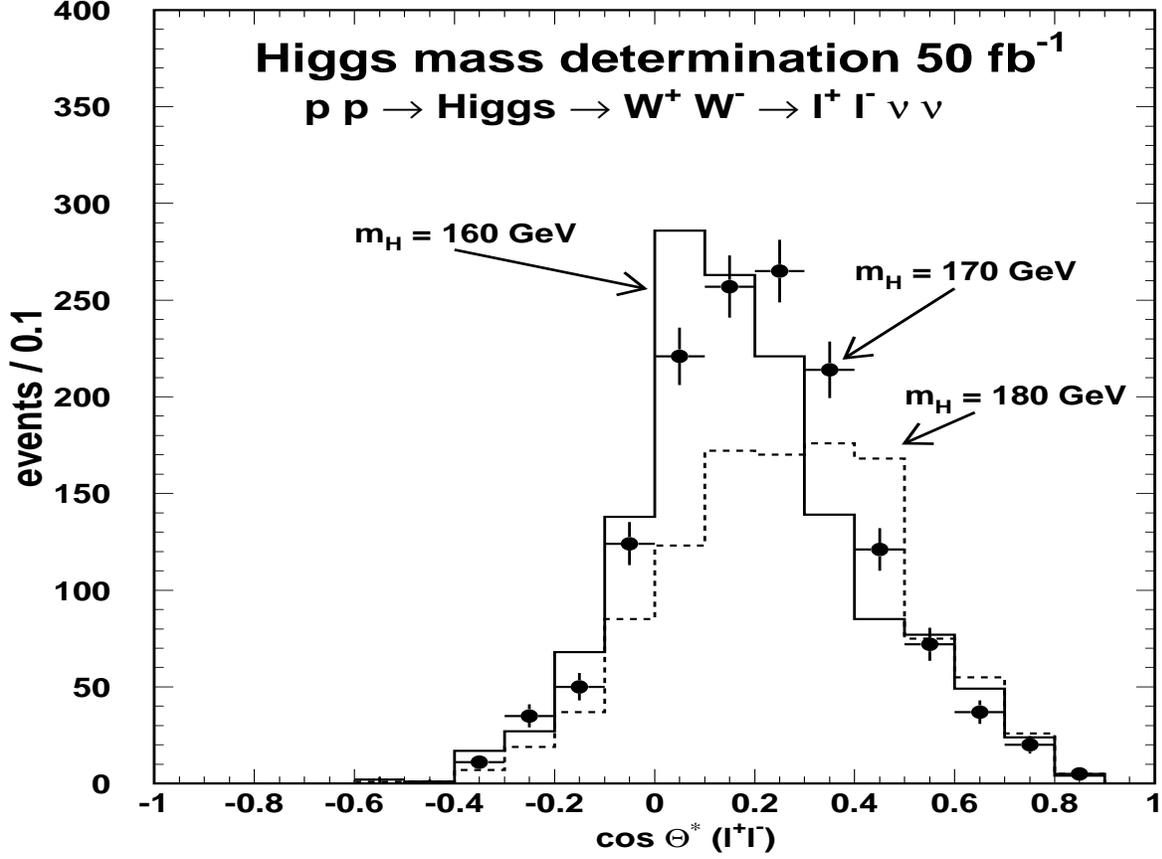}} 
   \vspace*{-1.9cm}
\caption[dummy]{\label{cosstar} \small Signal (shaded histogram) and background 
distribution of events in $\cos\theta_{\ell^+\ell^-}$ for $5\fbinv$ integrated 
luminosity.} 
   \vspace*{-0.5cm}
\end{figure}
For the dominant background process (\ref{eq:qqww}), the initial state is 
unpolarized. The two spin-1/2 quarks combine to a mixed state of a spin-1 
triplet and a spin-0 singlet and all final state $W$-polarization combinations
are allowed
\begin{equation}
q+{\bar q}\ra W^+_LW^-_L,W^+_TW^-_T,W^\pm_L W^\mp_T,
\label{eq:qqwwpol}
\end{equation}
where the quantisation axis is along the W-boson momenta. For the first two
final states, the lepton momenta are correlated as for the signal, since spin-2
is prohibited by angular momentum conservation. For the last state the $W$-spins
are not {\it anti}-correlated and thus ${\vec{{\bf p}}}_{e^+}$, ${\vec {{\bf p}}
}_{e^-}$ are {\it not} positively correlated, in contrast to the signal events.
The capability of distinguishing the signal from the background based on the $W
$-spin correlation therefore depends on the relative magnitude of the production
mechanisms (\ref{eq:qqwwpol}). In Figure 4 we show the respective differential
cross sections as a function of the invariant mass of the $WW$-system. We see
that for a large fraction of the background, where $M_{WW}<200\gev$, the
production of $W^\pm_L W^ \mp_T$ is of the same order as $W^+_TW^-_T$ and both
are much larger than $W^+_L W^-_L$. At $M_{WW}=165\gev$, $W^\pm_L W^\mp_T$ is
very close to half the production rate. We therefore expect a substantial
fraction of the background charged leptons to have a large relative opening
angle in contrast to the signal events. In Figure 5 we show the raw
distribution ({\it i.e.} before any other cuts) of $d\sigma/d\cos\theta_{\ell
\ell}$ for the signal and the background. Here $\theta_{\ell\ell }$ is the
relative angle between ${\vec{{\bf p}}}_{e^+}$ and ${\vec{{\bf p}}}_{e^-}$. The
signal almost vanishes for anti-parallel leptons,
$\cos\theta_{\ell^+\ell^-}=-1$, whereas the background has its maximum. We
impose the cut $10^o<\Delta\phi_\perp(\ell\ell)<45^o$ which is an extension of
cut {\it (vii)}. The effect is shown in Figure 2 of Ref\cite{paper}).

The combined effect of cuts ({\it viii}) and ({\it ix}) is $S/\sqrt{B_{WW}}
(M_{h^0})=12$ and $S/{B_{WW}}=1/1.3$, which is a substantial 
improvement. For the complete background we now have $S/\sqrt{B}8.4$ and 
$S/{B}=1/3$.

\subsection{Remaining Cuts}
The mass of the $WW$ events is given by $M_{WW}^2=(\sum_{\ell,\nu}E_i)^2-(\sum_
{\ell,\nu}{\vec{p}}_i)^2$. This can not be reconstructed because of the
neutrinos. However, based on the previous cuts, our signal events typically 
consist of two central, charged leptons with parallel momenta. Assuming $p_t
(h^0)\approx0$, we can thus approximate: $p_t(\nu\nu)\approx p_t(\ell\ell)$. 
In addition, we estimate the mass of the neutrinos to be equal to that of the 
charged leptons. Thus $E_{\nu\nu}\approx\sqrt{m_{\ell\ell}^2+p_t^2(\ell\ell)}$,
giving an overall estimate of $M^*_{WW}$ for $M_{WW}$. We expect this to be a 
good estimate for the signal events and require $M_{WW}^*>140\gev$. The 
distribution of the events as a function of $M_{WW}^*$ is shown in Figure 3, 
Ref.\cite{paper}. The background shows a much broader distribution in $M_{WW}^*
$, particularly at low values.

We define the variable $\theta^*$ as the opening angle between the lepton with
larger $p_t$ boosted to the dilepton rest frame and the momentum vector of the
dilepton system. The distribution of events in $\cos\theta^*$ is shown in
Figure 4 of Ref.\cite{paper} for $M_{h^0}=170\gev$. It shows a clear mass-peak
like behaviour which is significantly narrower than the background. We require
$0<\cos\theta^*<0.3$. We then obtain $S/\sqrt{B_{WW}}=12$ and $S/{B_{WW}}=2/1$,
which is tremendous. For the complete background we now have $S/\sqrt{B}=8$ and
$S/{B}1/1.2$. This is our final result. 

In Figure 6 we show how the peak in $\cos\theta^*$ shifts with Higgs mass. It is
clearly quite sensitive. Judging by this figure it looks possible to
determine the Higgs mass to within $\pm5\gev$ for an integrated luminosity of
$5\fbinv$. Again this compares very favourbaly with all other Higgs searches at
the LHC.

\section{Conclusion}
We have shown that the decay mode $h^0\ra W^+W^-\ra\ell^+\ell^-\nu\nu$ is an 
excellent Higgs search mode at the LHC for $M_{h^0}=155-180\gev$. A discovery
is possible for only $5\fbinv$. In addition, despite the absence of a
reconstructed narrow mass peak, the Higgs mass can be determined to
$\pm5\gev$ for $5\fbinv$. This is at least as good as any other Higgs
search mode. It thus more than fills the gap expected at $155-180\gev$ for the
first few years of LHC running. In our analysis, we have employed two new cuts
based on the boost and the spin-correlation of the $WW$-system. This enabled
the difficult separation from the irreducible continuum background production.

\end{document}